\documentclass[%
superscriptaddress,
amsmath,amssymb,
aps,
showkeys,
twocolumn,
]{revtex4-2}
\usepackage{hyperref}
\hypersetup{colorlinks,allcolors=blue}
\usepackage{graphicx}
\usepackage{bm}
\usepackage[all]{hypcap} 
\usepackage{color}
\usepackage{siunitx}
\usepackage{mathtools}
\usepackage{ulem}

\begin{document}

\title{Spin- and valley-dependent transports through ferromagnetic 8-pmmn borophene monolayer}

\author{Fatemeh Imanian Mofrad Bidgoli}
\affiliation{\footnotesize{Institute of Nanoscience and Nanotechnology, University of Kashan, Kashan 51167-87317, Iran}} 

\author{Hossein Nikoofard}
\affiliation{\footnotesize{Institute of Nanoscience and Nanotechnology, University of Kashan, Kashan 51167-87317, Iran}} 

\author{Narges Nikoofard}
\affiliation{\footnotesize{Institute of Nanoscience and Nanotechnology, University of Kashan, Kashan 51167-87317, Iran}} 

\author{Mahdi Esmaeilzadeh}
\email{mahdi@iust.ac.ir}
\affiliation{\footnotesize{Department of Physics, Iran University of Science and Technology, Tehran 16844, Iran}}

\date{\today}

\begin{abstract}
We study spin and valley-dependent transport properties in a n-p-n junction of 8-pmmn borophene monolayer. An external gate voltage and an exchange magnetic field, induced by the proximity effect of a ferromagnetic insulator, are applied to this junction as electric and magnetic potential barriers. We show that the exchange magnetic field generates spin polarization in the system and applying a gate voltage, as a simple method, causes valley polarization. This property (valley polarization) is due to the anisotropic and tilted Dirac cones of borophene structure and it is an advantage of borophene monolayer over graphene monolayer because in graphene it is necessary to apply strain in order to have valley polarization. We also show that the proposed device (borophene-based n-p-n junction) can work as perfect spin and perfect valley filters. The spin and valley filters can be controlled by changing  two factors, i.e., gate voltage and Fermi energy. Moreover, it is shown that for full spin and valley polarizations and thus perfect spin and valley filters, the length of the barriers must be larger than a specific value ($60$ $\mathrm{nm}$). These results show that borophene monolayer has a suitable potential to be used in spintronic and valleytronic devices.
\end{abstract}


\maketitle
\section{Introduction}
 
Discovering of two-dimensional (2D) materials are developing due to their interesting physical properties and promising potential for fabrication of electronic, spintronic devices, and integrated circuits \cite{huang2019,dery2011}. 2D materials with higher applications and beneficial attributes are of interest \cite{vishkayi,shukla}. For example, 2D materials with linear Dirac dispersion have significant transport properties due to the absence of backscattering, namely the Klein tunneling phenomena \cite{bib7}.

Among them, borophene, a monolayer of boron atoms, was introduced after its synthesis on the Ag surface \cite{shukla}. This 2D material has anisotropic electronic properties and excellent mechanical properties with a higher young’s modulus than that of graphene \cite{liu2018,xiang,wang}. Borophene as a substrate for sensing CO, NO, NO$_{2}$, and NH$_{3}$ gas molecules plays an important role in the development of 2D gas sensors \cite{tian}. It also has high flexibility and it is not easily oxidized \cite{xie}. The light mass of boron atoms causes borophene to be a promising material to be used in storing hydrogen and metal ion batteries \cite{wang,jena2017}. Besides, the very low lattice thermal conductivity, structural diversity and binding complexity introduce borophene as a new material for use in high-speed and low-dissipation electronic and spintronic devices \cite{thermal,norouzi,zeng}. A theoretical approach has shown that two allotropes of borophene, $\beta_{12}$ and $\delta_{6}$ can be used in nano-optoelectronics as a transparent conductor \cite{adamska2018}. Another allotrope, 8-pmmn borophene, with a unit cell including eight boron atoms, is a semiconductor with no band gaps and high mobility \cite{mannix2015,feng2016,zhong2017}. The 8-pmmn borophene has tilted Dirac cone than that of graphene that causes some specific properties such as asymmetric and oblique Klein tunneling \cite{oblique}.

Some properties of charge transport and conductance of borophene have been investigated by researchers. A study of magnetic properties with first-principle calculations indicates that the borophene nanoribbon with a zigzag edge has a ferromagnetic ground state \cite{zou2018}. The roles of tilted velocity and anisotropic Dirac cones on valley-dependent electron retroreflection and optical conductivity have been discussed in 8-pmmn borophene \cite{zhou3,verma2017}. From a theoretical approach, the effect of intrinsic line defects through a borophene monolayer on the charge transport is studied which identifies this material as a novel switching device \cite{zeng}. The effects of light on the charge conductance in irradiated 8-pmmn borophene monolayer have been also studied \cite{napitu}. Transmission probability and conductance in hydrogenated borophene have been studied and shown that it can be used as a filter for wavevector \cite{das2020}. In Ref. \cite{zheng2020}, it has been shown that if barrier orientation in 8-pmmn borophene be perpendicular to the tilted direction of Dirac cones, the valley polarization (or valley filtering) will be nearly 100$\%$. In this reference, to obtain valley filtering, both magnetic and electrical barriers must be applied to 8-pmmn borophene. Recently, Sattari has studied a 8-Pmmn borophene superlattice with electrostatic barriers. He was shown that number of the barriers, strength of the Rashba interaction and direction of superlattice are important parameters for adjusting the valley/spin polarization and conductance. Also, it is possible to obtain a valley polarization close to 100$\%$ under Rashba interaction {sattari1,sattari2}. 

Although the spin dependent transport properties have been studied vastly in famous 2D materials such as graphene, silicene, phosphorene, and TMDs \cite{yokoyama1,yokoyama2,Hedayati,zhang2016}, the spin dependent properties of 8-pmmn borophene sheet have not been studied. In the present work, spin and valley dependent transport properties in a n-p-n junction of 8-pmmn borophene monolayer are studied. An exchange magnetic field, induced by the ferromagnetic insulator, is applied to the system to obtain spin polarization and spin filter. We show that wave vector spin filtering is perfect and controllable by changing Fermi energy. Also, valley polarization and thus valley filtering can occur in this junction by using a simple method, i.e., by applying only a gate voltage. This is a prominence of borophene, because in the other famous 2D materials such as graphene, valley polarization is observed by using difficult methods such as applying a strain on graphene sheet \cite{Jalil,Katsnelson}. Similar to wave vector spin filter, wave vector valley filter is perfect and controllable by changing the Fermi energy. The spin and valley polarizations can be also controlled by tunning the gate voltage. This is a potential for borophene to be used as spin and valley field-effect transistors. Also, we show that full spin polarization and valley polarization occur when the length of the potential barriers is greater than a specific value i.e., $60$ $\mathrm{nm}$. According to these results, borophene monolayer can be considered as a suitable material for spintronic and valleytronic devices. 

The rest of the paper is organized as follows. In Sec. II, Hamiltonian and wave functions of the proposed device are presented and formulas for transmission probability and spin and valley polarization are obtained. In Sec. III, numerical results of the spin and valley transport properties and the influence of the system parameters are shown and discussed. Finally, a summary and conclusion are given in Sec. IV. 

\section{THEORETICAL MODEL} \label{sec2}

We consider a n-p-n junction based on borophene monolayer as shown in Fig. 1. To induce and control spin and valley resolved electron transport, an exchange magnetic field induced by proximity effect of a ferromagnetic insulator and an external gate voltage is applied in the central region with length $L$ (see Fig. 1). The effective Hamiltonian of this system in a tight-binding model is given by \cite{zhou3}

\begin{equation}
H=\xi \hbar(v_x\sigma_x k_x+v_y\sigma_y k_y+v_tk_y\mathbf{I})+(eV_{g}+\sigma M)\mathbf{I},
\end{equation}
where $\xi=\pm 1$ is the valley index for the valleys $K$ and $K'$, respectively, $\hbar$ is the Plank constant, $v_x=0.86 v_F$ and $v_y=0.69 v_F$ are anisotropic velocities, and $v_t=0.32 v_F$ is tilted velocity in which $v_F=10^6$ $\mathrm{m/s}$ is the Fermi velocity. In Hamiltonian equation, $\sigma_x$ and $\sigma_y$ are the set of Pauli matrices, $\mathbf{I}$ is the unit matrix, $k_x$ and $k_y$ are the $x$ and $y$ components of the wavevector for the incoming (outgoing) wave to (from) the barrier (p or central region, see Fig. 1), $-e$ is the electron charge, $V_{g}$ is the gate voltage, $M$ is the normalized strength of exchange magnetic field, and $\sigma=\pm 1$ is the spin index. 

Diagonalizing the Hamiltonian, the eigenvalues of the system for conduction and valence bands can be written as
\begin{equation}
E=eV_{g}+\sigma M+\xi \hbar v_tk_y\pm \hbar \sqrt{(v_xk_x)^2+(v_yk_y)^2},
\end{equation}
where sign $+ (-)$ in Eq. (2) denotes the conduction (valence) band. 

In device shown in Fig. 1, we introduce three regions: left (L) and right (R) regions and a central (C) region with length $L$ which is perturbed by a gate voltage and an exchange magnetic field. We confine the central region with boundaries $x_1$ and $x_2$ as shown in the figure. One can consider electrons incident to the central region with a specific value of $k_{y}$ and a given value of Fermi energy $E$. Using Eq. (2), $k_{x}$ can be obtained as 
\begin{equation}
k_x=\pm \frac{1}{\hbar v_x}\sqrt{(E-\xi \hbar v_tk_y)^2-(\hbar v_yk_y)^2},
\end{equation}
where plus sign denotes the incident and transmitted electrons with wave vector $k^{i}_x$ and $k^{t}_x$ in the left and right region, respectively and minus sign indicates the reflected electrons with wave vector $k^{r}_x$ in the left region. Similar calculation shows that the wave vector of electrons in the central region is given by
\begin{equation}
q_x=\pm \frac{1}{\hbar v_x}\sqrt{(E-eV_{g}-\sigma M-\xi \hbar v_tk_y)^2-(\hbar v_yk_y)^2}.
\end{equation}

Here same as left and right regions, $+ (-)$ sign denotes the incident (reflected) electrons with wave vector $q^{i(r)}_x$.

The wave functions of Hamiltonian in Eq. (1) for the three regions of device can be written as \cite{zhou3}
\begin{equation}
\begin{aligned}
& \Psi_L=\binom{1}{A_i}e^{i(k_{x}^ix+k_{y}y)} \\
& \qquad +r_{\xi}^{\sigma}\binom{1}{A_r}e^{i(k_{x}^rx+k_{y}y)}    \qquad   x<x_{1},
\end{aligned}
\end{equation}
\begin{equation}
\begin{aligned}
& \Psi_C=a_1\binom{1}{B_i}e^{i(q_{x}^ix+k_{y}y)} \\
& \qquad +b_1\binom{1}{B_r}e^{i(q_{x}^rx+k_{y}y)}      \qquad     x_{1}<x<x_{2},
\end{aligned}
\end{equation}
\begin{equation}
\Psi_R=t_{\xi}^{\sigma}\binom{1}{A_i}e^{i(k_{x}^tx+k_{y}y)}   \qquad                      x>x_{2},
\end{equation}
where 
\begin{equation}
\begin{aligned}
&  A_{i(r)}=\xi \hbar \frac{v_xk^{i(r)}_x+iv_yk^{i(r)}_y}{E-\xi \hbar v_tk^{i(r)}_y},  \\
&  B_{i(r)}=\xi \hbar \frac{v_xq^{i(r)}_x+iv_yq^{i(r)}_y}{E-\xi \hbar v_tq^{i(r)}_y-eV_{g}-\sigma M},
\end{aligned}
\end{equation}
$r_{\xi}^{\sigma}$ and $t_{\xi}^{\sigma}$ are respectively the spin and valley dependent reflection and transmission coefficients, and $a_1$ and $b_1$ are scattering amplitudes in the barrier region. Here $k_y$ is preserved throughout the system due to invariance along this direction. Using the above wave functions and applying the boundary conditions at $x_{1}$ and $x_{2}$, the reflection and the transmission coefficients can be obtained for the incident wave from the left side as follows

\begin{equation}
r_{\xi}^{\sigma}=\frac{(A_i-B_i)(B_r-A_i)(e^{iq_{x}^i (x_{2}-x_{1})}-e^{iq_{x}^r (x_{2}-x_{1})})}{D},
\end{equation}
\begin{equation}
t_{\xi}^{\sigma}=\frac{(A_i-A_r)(B_i-B_r)e^{i(q_{x}^i+q_{x}^r-k_{x}^i)(x_{2}-x_{1})}}{D},
\end{equation}
where
\begin{equation}
\begin{aligned}
& D=(A_i-B_i)(A_r-B_r)e^{iq_{x}^i (x_{2}-x_{1})} \\
& \qquad -(B_r-A_i)(B_i-A_r)e^{iq_{x}^r (x_{2}-x_{1})}.
\end{aligned}
\end{equation}
\begin{figure}[!h]\label{fig1} 
\centering
\includegraphics[scale=0.32]{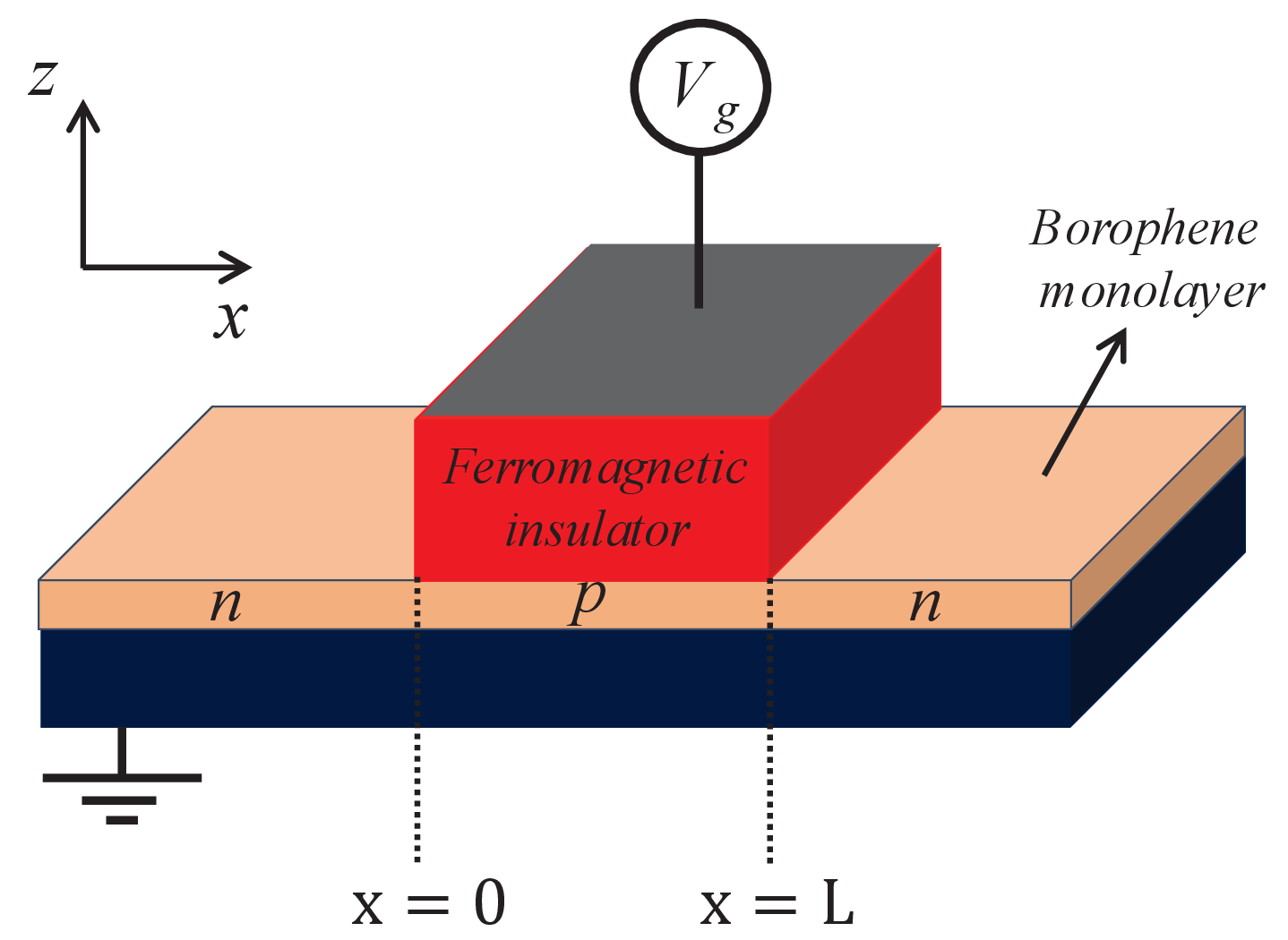}
\caption{Schematic view of a n-p-n junction based on a borophene monolayer. An exchange magnetic field, induced by the proximity effect of a ferromagnetic insulator, and a gate voltage $V_g$ are applied in the central region with length $L$. We also use an insulator (shown in dark blue color) for excluding the permeate of the gate current.}\label{fig1}
\end{figure}

Eq. (10) gives the transmission probability of electron through the device as $T_{\xi}^{\sigma}=|t_{\xi}^{\sigma}|^2$. 
The spin-dependent transmission probability for spin-up (${\uparrow}$) and spin-down (${\downarrow}$) are
\begin{equation} \label{myeq1}
T^{\uparrow (\downarrow)}=T_{K}^{\uparrow (\downarrow)}+T_{K'}^{\uparrow (\downarrow)}.
\end{equation}

Also, the valley-dependent transmission probability for valleys $K$ and $K'$ are given by
\begin{equation} \label{myeq1}
T_{K(K')}=T_{K(K')}^{\uparrow}+T_{K(K')}^{\downarrow}.
\end{equation}

The spin and valley polarization are defined by 
\begin{equation} 
P_s=\frac{T^{\uparrow}-T^{\downarrow}}{T^{\uparrow}+T^{\downarrow}},
\end{equation}
and 
\begin{equation} 
P_v=\frac{T_{K}-T_{K'}}{T_{K}+T_{K'}},
\end{equation}
respectively. Here $P_s = \pm {1}$ ($P_v = \pm {1}$) indicates the fully spin (valley) polarized situation.

\section{RESULTS AND DISCUSSION}\label{sec3}
     
To study the transport properties of system numerically, we consider a n-p-n junction of 8-pmmn borophene monolayer in which an exchange magnetic field is applied in the central region (see Fig. 1). The exchange field is considered to be perpendicular to borophene layer i.e., in the $z$ direction with normalized strength $M=80$ $\mathrm{meV}$ \cite{norouzi}. To control the electron transport, an external gate voltage is also applied in central region. 
\begin{figure*} 
\centering
\includegraphics[scale=0.8]{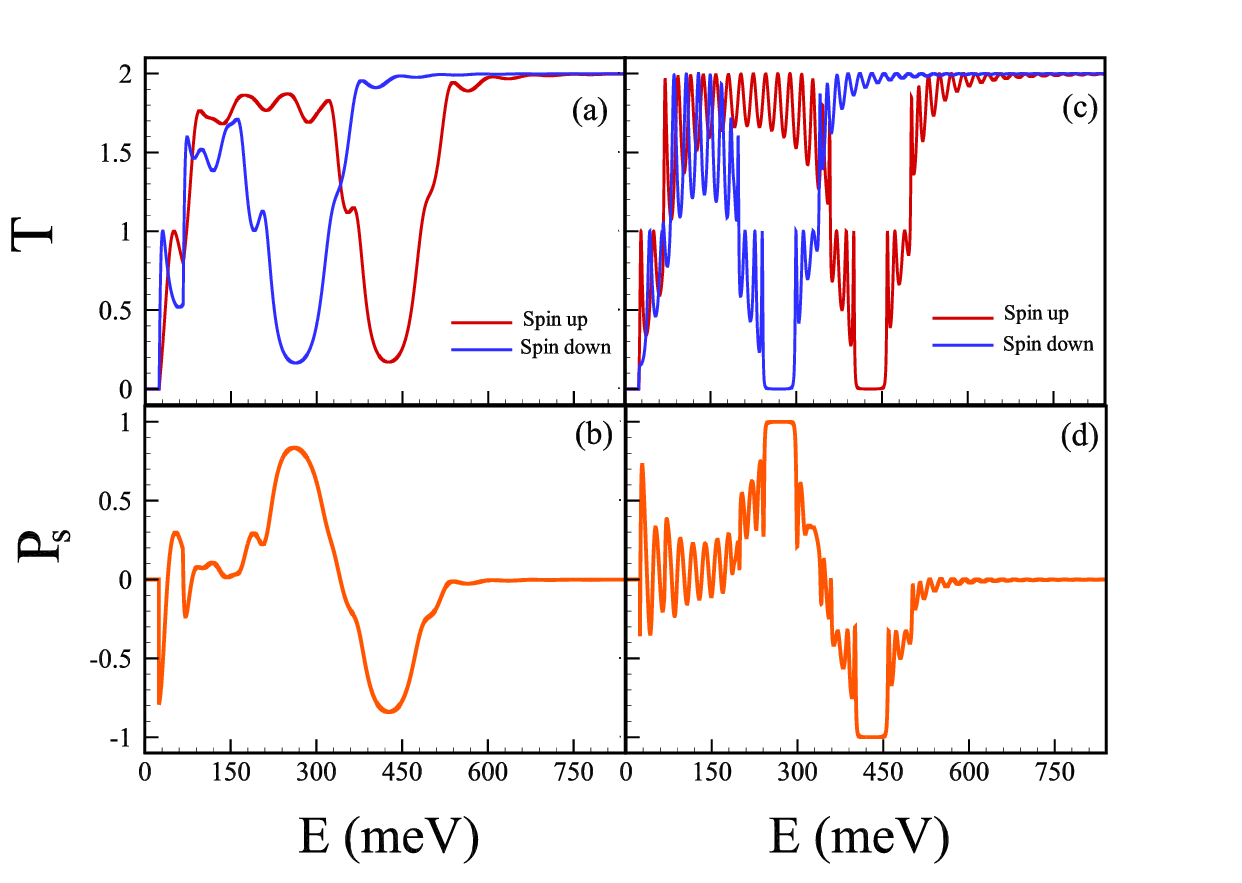}
\caption{(a) Spin-dependent transmission probability and (b) spin polarization versus the Fermi energy $E$ for $L=25$ $\mathrm{nm}$. (c) and (d) are respectively the same as (a) and (b) but for $L=80$ $\mathrm{nm}$. Other parameters are $V_{g}=350$ $\mathrm{meV}$, $k_y=0.1$ $\mathrm{nm^{-1}}$, and $M=80$ $\mathrm{meV}$.}
\end{figure*}
\begin{figure}[!h]
\centering
\includegraphics[scale=0.45]{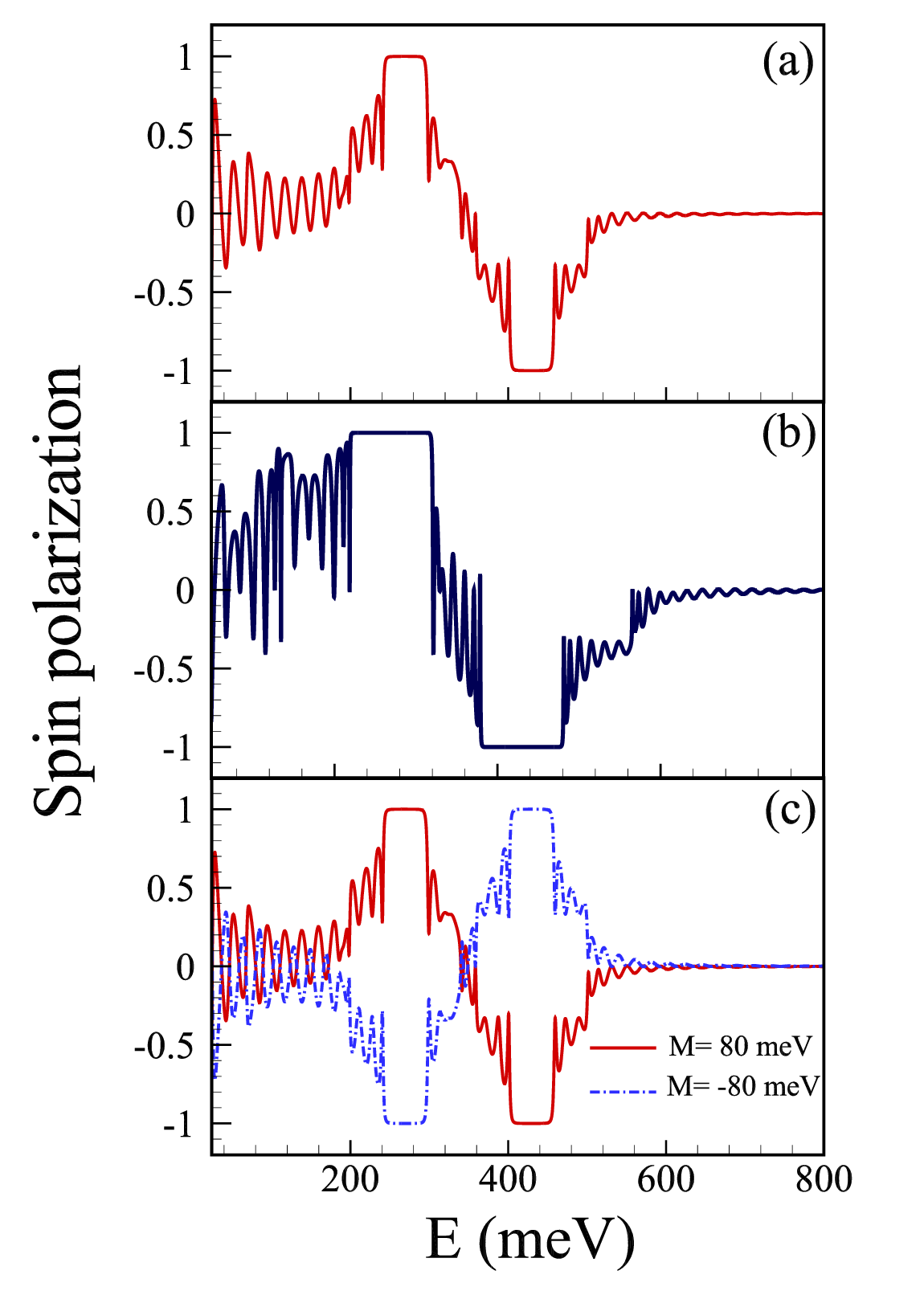}
\caption{Spin polarization versus the Fermi energy $E$ for wave vectors (a) $k_y=0.1$ $\mathrm{nm^{-1}}$ and (b) $k_y=0.2$ $\mathrm{nm^{-1}}$. (c) Spin polarization versus the Fermi energy $E$ for two values of exchange field: $M=80$ $\mathrm{meV}$ and $M=-80$ $\mathrm{meV}$. Other parameters are $V_{g}=350$ $\mathrm{meV}$ and $L=80$ $\mathrm{nm}$.}
\end{figure}

Figure 2 shows the spin-dependent transmission probability and spin polarization versus the Fermi energy for two lengths of barrier (central) region, i.e., $L=25$ $\mathrm{nm}$ and $L=80$ $\mathrm{nm}$. As seen in Figs. 2(a)-2(d), the exchange magnetic field breaks the spin degeneracy and causes the transmitted electrons to be spin-polarized. In Figs. 2(a) and 2(c), the oscillating behavior of transmission probability is related to the interference effect between the incident and the reflected electronic waves in the barrier region. Also, the number of fluctuations increases with increasing barrier length $L$. This oscillatory behavior is an evidence for the existence of Dirac-like quasiparticles in borophene which is in contrast to the exponential nature of Schrodinger quasiparticles in other materials \cite{das2020}. Regarding to the real values of wave vector for incident electron, $k_x$, we consider an allowable range of Fermi energies for electrons to flow throughout the system (see Eq. (A.2) of Appendix for more details). 

In Fig. 2(c), we see an interval of energy with zero transmission probability for spin up (i.e., $240<E< 300$ $\mathrm{meV}$) and spin down (i.e., $400<E< 460$ $\mathrm{meV}$). This is due to the imaginary value of the wave vector in the potential region (i.e., imaginary value of $q_x$) that leads to the participation of evanescence mode in transport \cite{napitu}. These intervals of energy can be calculated simply by considering Eq. (A.3) of the Appendix. For other values of Fermi energy, propagation modes participates in transport. In Figs. 2(a) and 2(c), the maximum value of transmission probability is equal to $2$. It is because according to Eq. (12),  spin-dependent transmission is the summation of two valleys' transmission probabilities with a maximum value of $1$ for each valley.

A comparison between Figs. 2(b) and 2(d) shows increasing the value of barrier length $L$ from $25$ $\mathrm{nm}$ to $80$ $\mathrm{nm}$, we have perfect spin polarization, i.e., $P_s = \pm{1}$. In other words, for $L=25$ $\mathrm{nm}$, as shown in Fig. 2(b), partial spin polarization occurs with maximum (minimum) equals to $+0.75$ ($-0.75$). While, for $L=80$ $\mathrm{nm}$, as shown in Fig. 2(d), complete spin polarization occurs with maximum (minimum) equals to $+1$ ($-1$). As we will show later the perfect spin polarization can be achieved for $L>60$ $\mathrm{nm}$. In Fig. 2(d), the spin polarization $P_s$ changes between $-1$ and $+1$. So, we can control the spin polarization by tuning the Fermi energy. The Fermi energy can be tunned via a back gate \cite{prada2009}. Also, the proposed device can work as a spin-up filter with $P_s = +1$ in the energy interval $240<E< 300$ $\mathrm{meV}$, and a spin-down filter with $P_s = -1$ in the energy interval $400<E<460$ $\mathrm{meV}$. For $E>700$ $\mathrm{meV}$, $P_s$ goes to zero which shows unpolarized transmission.

Figure 3 shows the effects of $k_y$ and changing the direction of magnetization on the spin polarization. As shown in Figs. 3(a) and 3(b), we observe that the increase of wave vector in $y$-direction from $k_y=0.1$ $\mathrm{nm^{-1}}$ to $k_y=0.2$ $\mathrm{nm^{-1}}$ causes the energy intervals for perfect polarization (gap regions) become larger. We know that the transmission probability in gap regions has a zero value for one spin index and a nonzero value for the other spin index. According to Eq. (A.4) of Appendix, the gap region depends on the $k_y$. Therefore, we can increase the energy regions in which perfect spin polarization (or spin filtering) can take place by increasing $k_y$. In Fig. 3(c), we see that by changing the direction of magnetization from ($-z$) to ($+z$), the sign of spin polarization changes from $+1$ to $-1$ and vice versa as expected. It is worth to mention that changing the magnetization direction is not easy same as changing the electron Fermi energy (by back gate) as mentioned before.

\begin{figure}[!h]
\centering
\includegraphics[scale=0.45]{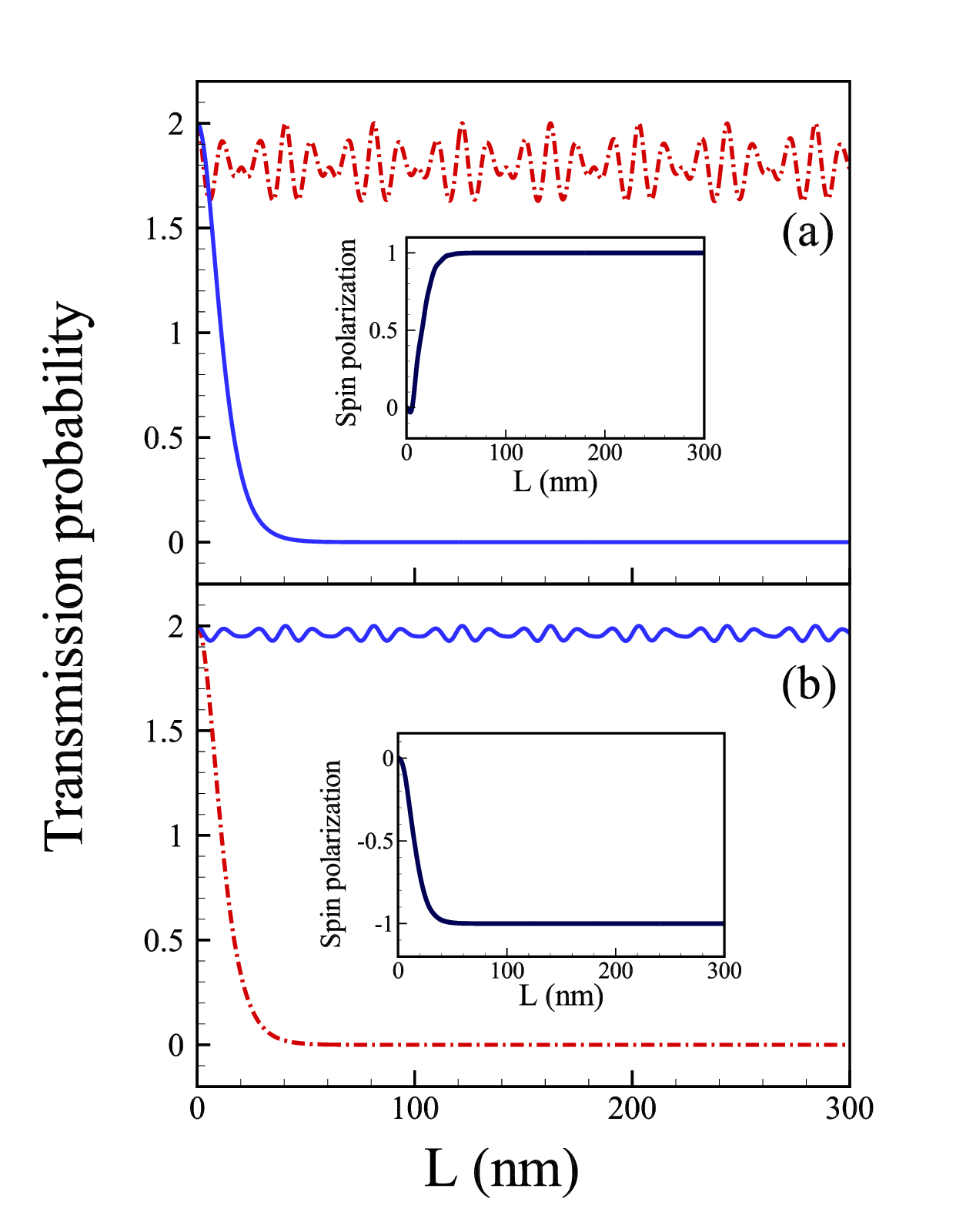}
\caption{Spin-dependent transmission probability versus the length of potential barrier region for two values of Fermi energy: (a) $E= 270$ $\mathrm{meV}$ and (b) $E= 430$ $\mathrm{meV}$. Red dashed (blue solid) line are related to to spin-up (spin-down) transmission probability. Insets show spin polarization. Other parameters are $V_{g}=350$ $\mathrm{meV}$, $k_y=0.1$ $\mathrm{nm^{-1}}$, and $M=80$ $\mathrm{meV}$.}
\end{figure} 
\begin{figure}[!h]\label{fig5} 
\centering
\includegraphics[scale=0.45]{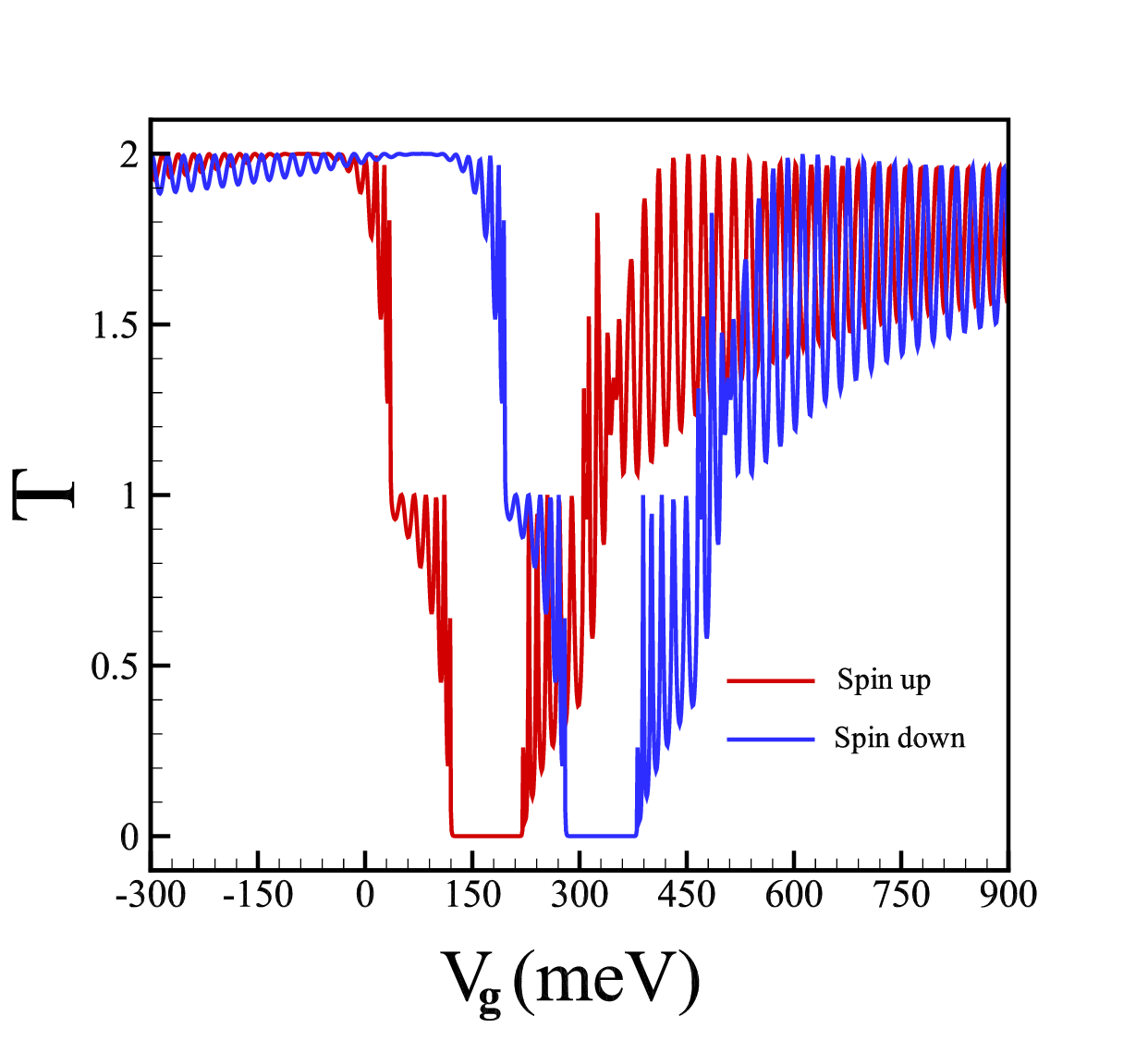}
\caption{Spin-dependent transmission probability versus the gate voltage for $E=250$ $\mathrm{meV}$. Other parameters are $k_y=0.2$ $\mathrm{nm^{-1}}$, $L=80$ $\mathrm{nm}$, and $M=80$ $\mathrm{meV}$.}
\label{fig9}
\end{figure}
In Fig. 4, we plot the spin-dependent transmission probability and spin polarization versus the length of the potential barrier (central region) for two special values of Fermi energy, i.e., $E= 270$ $\mathrm{meV}$ and $E= 430$ $\mathrm{meV}$. For $E= 270$ $\mathrm{meV}$, as indicated in Fig. 4(a), the spin-up transmission probability (red line) has oscillating behavior, but spin-down transmission probability (blue line) decays with increasing $L$ and goes to zero value for $L>60$ $\mathrm{nm}$ in which $P_s = +1$ as shown in the inset of figure. So, for $L>60$ $\mathrm{nm}$, the system can work as a perfect spin-up filter. Inversely, for $E= 430$ $\mathrm{meV}$, as we see in Fig. 4(b), the spin-down transmission probability (blue line) has an oscillating behavior, while spin-up transmission probability (red line) decays and goes to zero for $L>60$ $\mathrm{nm}$. Here for $L>60$ $\mathrm{nm}$, $P_s = -1$ [see the inset of Fig. 4(b)] and the system can work as a perfect spin-down filter. Therefore, the spin filtering is perfect for both spin up and spin down when the width of potential region exceeds a certain value ($60$ $\mathrm{nm}$). Also, spin filtering is controllable because one can change spin up filter to spin down filter and vice versa by changing the Fermi energy via a back gate. 
\begin{figure}[!h]\label{fig3} 
\centering
\includegraphics[scale=0.45]{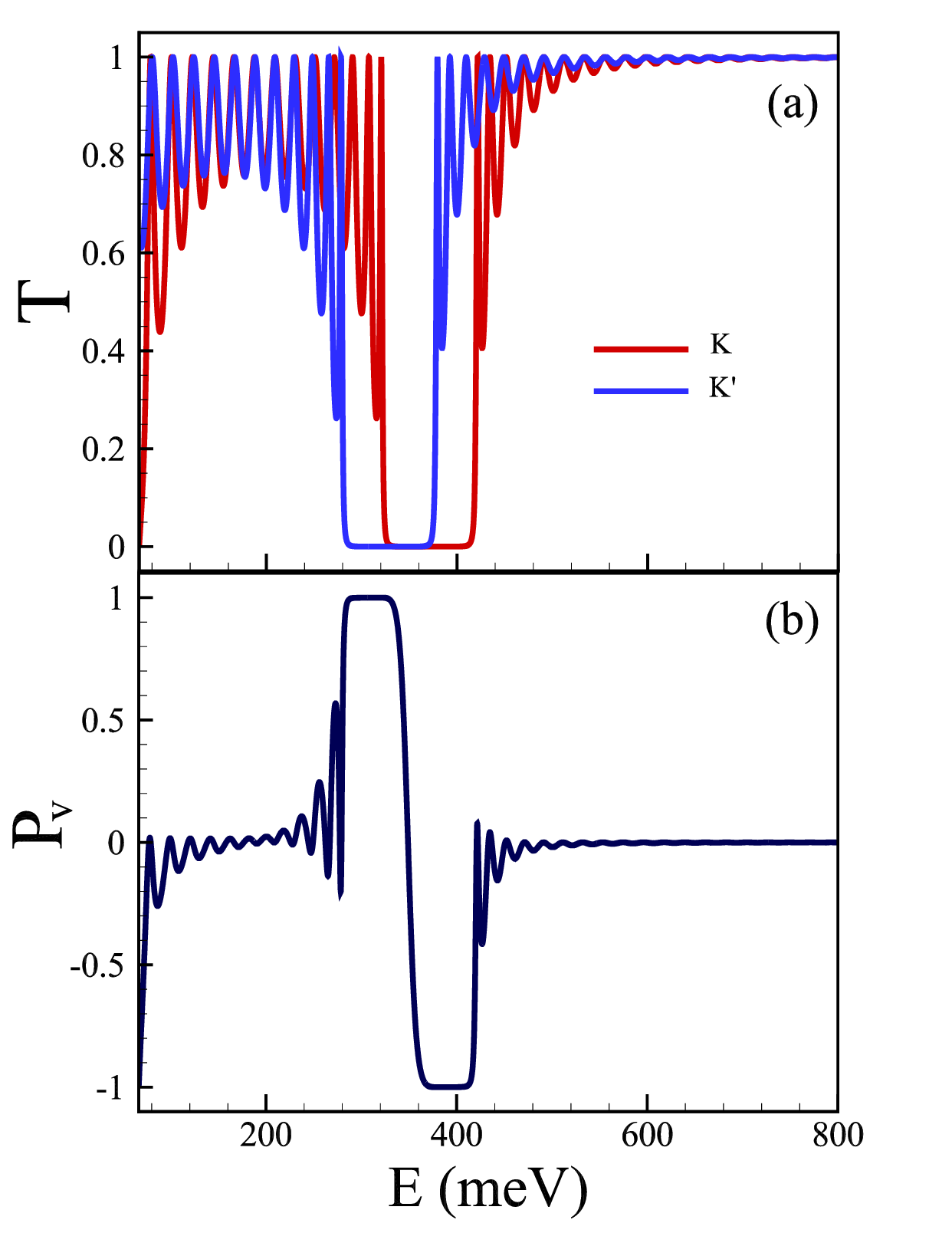}
\caption{(a) Valley-dependent transmission probability and (b) valley polarization versus the Fermi energy $E$ for gate voltage $350$ $\mathrm{meV}$ and in the absence of exchange magnetic field ($M=0$). Other parameters are $L=80$ $\mathrm{nm}$ and $k_y=0.1$ $\mathrm{nm^{-1}}$.}  
\label{fig3}
\end{figure}

Effect of gate voltage on spin-dependent transmission properties is shown in Fig. 5. It is observed that in the gate voltage interval $120<V_g< 220$ $\mathrm{meV}$, only electrons with spin down can pass through the device, while in the gate voltage interval $280<V_g< 380$ $\mathrm{meV}$, only electrons with spin up can pass. Thus, in intervals $120<V_g< 220$ $\mathrm{meV}$ and $280<V_g< 380$ $\mathrm{meV}$, the device can act as perfect spin down and spin up filters, respectively. As a result, one can change the spin direction of spin filter from down to up and vice versa by changing the gate voltage e.g., from $150 (350)$ to $350 (150)$ $\mathrm{meV}$. Therefore, in addition to using Fermi energy for changing of spin direction of spin filter, we can use gate voltage for this purpose.

Now, we study valley-dependent transmission properties and the results are shown in Figs. 6 and 7. Figure 6 shows valley-dependent transmission probability and valley polarization versus the Fermi energy in the absence of the exchange field ($M = 0$). As seen in Fig. 6(a), valley splitting can occur by applying only a gate voltage. This property is due to the anisotropic and tilted Dirac cones in the borophene structure \cite{Islam}. It is an important result that one can have valley polarized electron transport by using a simple way (applying only a gate voltage). This property is not observed for graphene as we will show later. Similar to Fig. 2, the gap regions in this figure are related to the evanescence modes of electronic waves in which wave vectors $k_x$ and $q_x$ have imaginary values, respectively \cite{napitu,das2020}. As we can see in Fig. 6(b), for intervals $280<E< 340$ $\mathrm{meV}$ and $360<E< 420$ $\mathrm{meV}$, the valley polarization is perfect i.e., $P_v = +1$ and $-1$, respectively. Thus the proposed device can work as perfect and controllable valley filter in which by changing Fermi energy one can change the valley polarization from $+1 (-1)$ to $-1 (+1)$. Our numerical calculations indicate that, same as perfect spin polarization (see Fig. 4), perfect valley polarization occurs for the lenght of potential barrier greater than a certain value ($L= 60$ $\mathrm{nm}$). 

\begin{figure}[!h]\label{fig6} 
\centering
\includegraphics[scale=0.45]{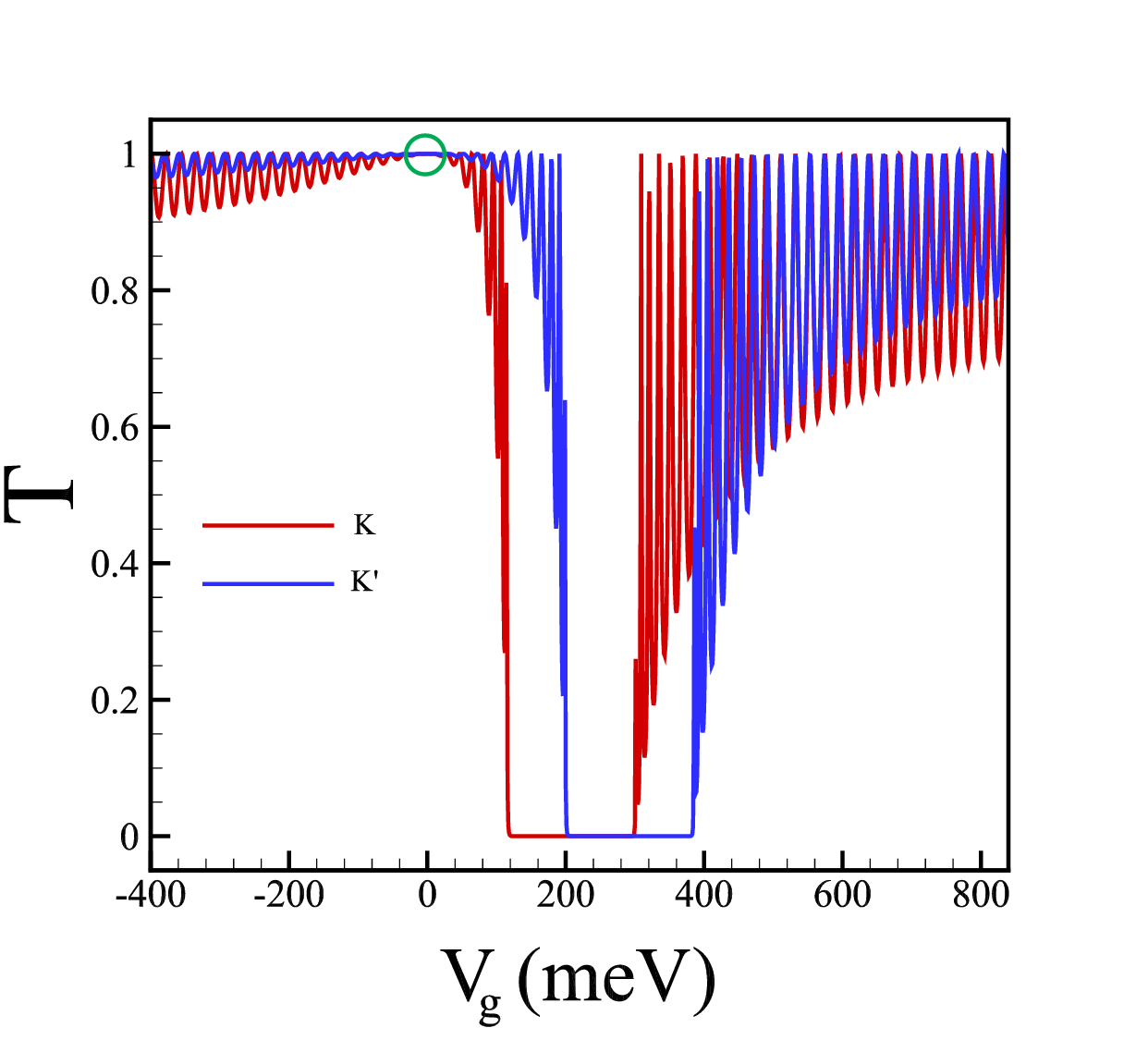}
\caption{Valley-dependent transmission probability versus the gate voltage for $E=250$ $\mathrm{meV}$. Other parameters are $k_y=0.2$ $\mathrm{nm^{-1}}$, $L=80$ $\mathrm{nm}$, and $M=0$. The green circle shows that near $V_g=0$, the transmission of electrons with $K$ and $K'$ valleys are equal which indicates unpolarized transmission.}
\label{fig9}
\end{figure}
Valley-dependent transmission probability versus the gate voltage for a fixed value of Fermi energy is shown in Fig. 7. It is observed that for gate potential interval $120<V_g< 200$ $\mathrm{meV}$, only electrons with valley $K'$ can transmit through the system, but for interval $300<V_g< 380$ $\mathrm{meV}$ only electrons with valley $K$ can transmit through the system. Thus, for the first and second intervals, the system can work as $K'$ and $K$ perfect valley filters, respectively.  Also, for interval $200<V_g< 300$ $\mathrm{meV}$, transmission probability for both valley $K$ and $K'$ is zero, therefore the system is off for both valleys. Moreover, near $V_g = 0$ (shown by a green circle in the figure), transmission probability for both valleys are equal which indicates unpolarized valley transmission. Consequently, by changing only the gate voltage, the system can work with four different modes: 1- as valley $K$ perfect filter in which only electrons with valley $K$ are allowed to pass, 2- as valley $K'$ perfect filter in which only electrons with valley $K'$ are allowed to pass, 3- as unpolarizer in which electrons with both valley $K$ and valley $K'$ are allowed to pass, and 4- off mode in which no electrons are allowed to pass.

\begin{figure}[!h]
\centering
\includegraphics[scale=0.45]{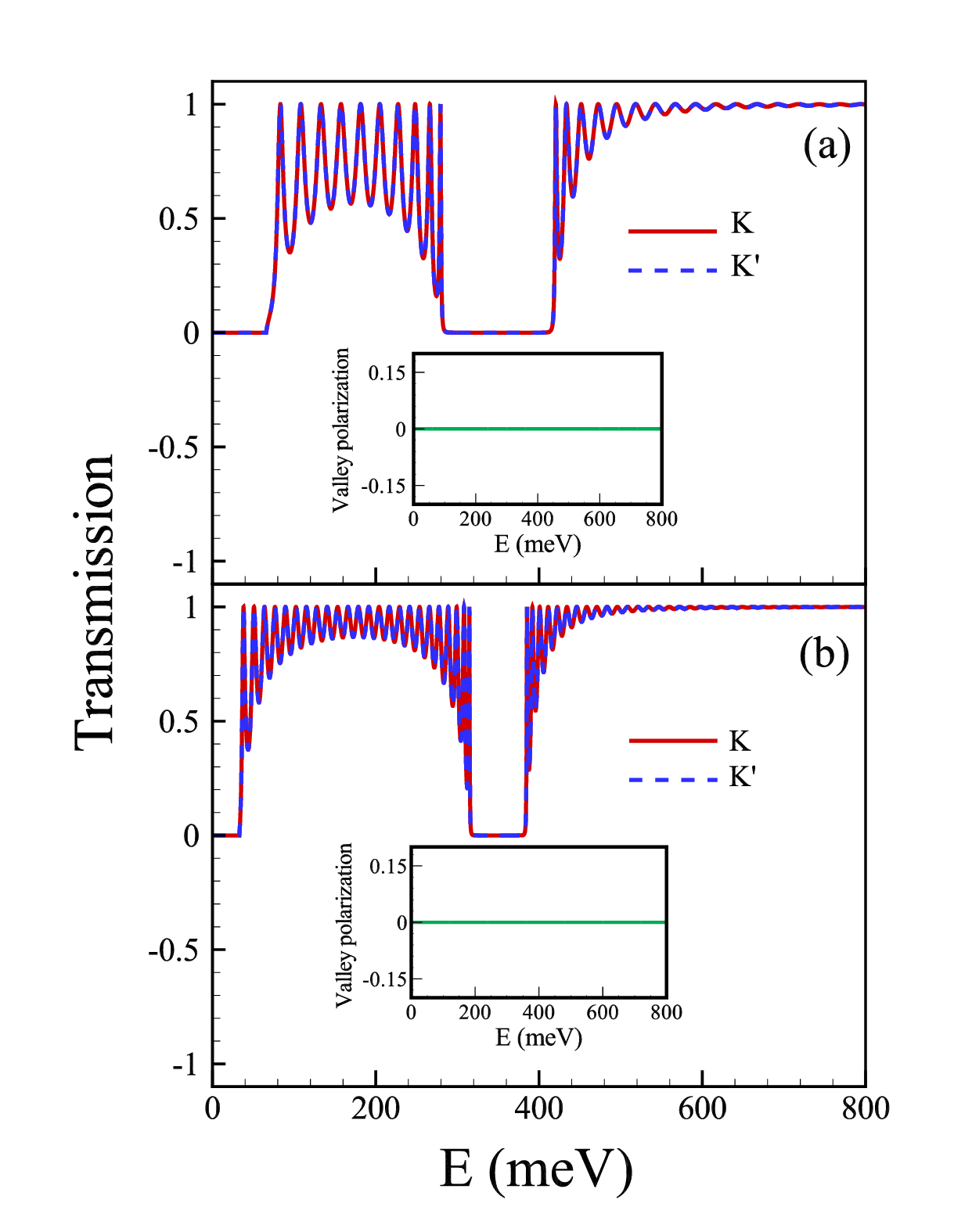}
\caption{Valley-dependent transmission probability versus the Fermi energy $E$ for (a) graphene and (b) silicene. The insets indicate valley polarization. This figure is plotted for the same conditions as used in Fig. 6.}
\end{figure}

At the end of this section, we compare valley-dependent transmission properties in borophene with two other important 2D materials i.e., graphene and silicene. Figure 8 shows valley-dependent transmission probability for these two materials at the same conditions of Fig. 6. As we can see in this figure, for both graphene and silicene, transmission probability of valley $K$ and valley $K'$ are completely the same which shows valley splitting and thus valley polarization cannot occur in these materials by applying only a gate voltage. This is the advantage of borophene relative to graphene and silicene which is due to its tilted and anisotropic Dirac cone. To obtain valley splitting in graphene and silicene, other methods such as applying strain should be used \cite{Jalil,Katsnelson}, which is not so simple.

\section{CONCLUSION}\label{sec4}

In this paper, we studied the spin and valley dependent transport properties in 8-pmmn monolayer of borophene. For this purpose, a n-p-n junction is considered in the presence of electric and magnetic barriers, i.e., an external gate voltage and an exchange magnetic field induced by the proximity effect of a ferromagnetic insulator. We showed that applying the exchange magnetic field causes spin polarization in which the sign of spin polarization changes with changing the direction of this field. Because the spin polarization is equal to $\pm{1}$, the proposed device can work as a perfect spin filter. The spin filter is controllable by two methods: 1- changing Fermi energy and 2- changing gate voltage. As a simple method, applying a gate voltage causes the valley polarization and valley filter in borophene. While in graphene, one needs complicated methods such as applying strain for valley polarization. This shows the advantage of borophene relative to graphene in valley polarization. As an interesting result, the spin and valley polarization can be controlled by changing two factors, i.e., gate voltage and Fermi energy. Furthermore, we showed that perfect spin and valley filters can occur for length of potential barriers greater than $60$ $\mathrm{nm}$. Our results could motivate researchers to consider borophene monolayer as a promising candidate for fabricating spin and valley filters in future spintronic and valleytronic industries.

\section*{Declaration of Competing Interest}
The authors declare that they have no known competing financial interests or personal relationships that could have appeared to influence the work reported in this paper.

\section*{Acknowledgment}
This work has not been supported by no foundation.

\section*{Appendix}
To obtain the allowable range of energy in which $k_x$ be real, we use Eq. (3). In this equation, below the radical must be positive, so
\begin{equation}\tag{A.1}
E^2-2\xi \hbar v_tk_yE+(\hbar k_y)^2(v_t^2-v_y^2)>0.
\end{equation}

Considering the condition of positivity of a quadratic polynomial, the energy intervals for a defined $k_y$ must be as follows
\begin{equation}\tag{A.2}
\begin{aligned}
& E>\hbar k_y(\xi v_t+v_y), \\
& E<\hbar k_y(\xi v_t-v_y).
\end{aligned}
\end{equation}

Substituting of constants parameters from section II and considering $k_y=0.1$ $\mathrm{nm^{-1}}$, intervals $E>24.3$ $\mathrm{meV}$ and $E>66.5$ $\mathrm{meV}$ are allowable for valleys $K'$ and $K$, respectively. Similarly, we use Eq. (4) to calculate the allowable range of energy in which $q_x$ is real. So, we have
\begin{equation}\tag{A.3}
\begin{aligned}
& E>eV_{g}+\sigma M+\hbar k_y(\xi v_t+v_y), \\
& E<eV_{g}+\sigma M+\hbar k_y(\xi v_t-v_y), 
\end{aligned}
\end{equation}

With attention to above equation, the gap region can be obtain as follows
\begin{equation}\tag{A.4}
\delta{E}=2\hbar k_{y}v_{y}.
\end{equation}

\end{document}